\documentclass[conference]{IEEEtran}
\IEEEoverridecommandlockouts

\usepackage{cite}
\usepackage{amsmath,amssymb,amsfonts}
\usepackage{algorithmic}
\usepackage{graphicx}
\usepackage{textcomp}
\usepackage{xcolor}
\usepackage{makecell}
\usepackage{hyperref}
\usepackage{listings}
\usepackage{caption}
\usepackage{pifont}
\usepackage{multirow}

 \def\cameraready{1111}  

\def\BibTeX{{\rm B\kern-.05em{\sc i\kern-.025em b}\kern-.08em
    T\kern-.1667em\lower.7ex\hbox{E}\kern-.125emX}}
\begin{document}

\newcommand\todo[1]{{\color{red}(TODO: #1)}}

\title{GSIM: Accelerating RTL Simulation for Large-Scale Designs}

\ifx\cameraready\undefined
\else

\author{\IEEEauthorblockN{
Lu Chen$^{*\dag}$,
Dingyi Zhao$^\ddag$,
Zihao Yu$^*$,
Ninghui Sun$^{*\dag}$ and 
Yungang Bao$^{*\dag}$}
\IEEEauthorblockA{
$^*$\textit{State Key Lab of Processors, Institute of Computing Technology, Chinese Academy of Sciences, China} \\
$^\dag$\textit{University of Chinese Academy of Sciences, China} \\
$^\ddag$\textit{Beijing Institute of Open Source Chip, China} \\
}}

\fi

\maketitle

\begin{abstract}
Register Transfer Level (RTL) simulation is widely used in design space exploration, verification, debugging, and preliminary performance evaluation for hardware design. Among various RTL simulation approaches, software simulation is the most commonly used due to its flexibility, low cost, and ease of debugging. However, the slow simulation of complex designs has become the bottleneck in design flow. In this work, we explore the sources of computation overhead of RTL simulation and conclude them into four factors. To optimize these factors, we propose several techniques at the supernode level, node level, and bit level. Finally, we implement these techniques in a novel RTL simulator GSIM. GSIM succeeds in simulating XiangShan, the state-of-the-art open-source RISC-V processor. Besides, compared to Verilator, GSIM can achieve speedup of 7.34x for booting Linux on XiangShan, and 19.94x for running CoreMark on Rocket.


\end{abstract}


\section{Introduction}


Register Transfer Level (RTL) simulation is essential in hardware design. It is widely used in various aspects, including design space exploration, verification, debugging, and preliminary performance evaluation~\cite{repcut} \cite{khronos} \cite{essent}. It has become an indispensable step in the chip development workflow.


There are currently three main approaches to RTL simulation: software simulation, FPGA prototyping, and hardware emulation. Among them, FPGA prototyping is the fastest, but debugging tools are limited. Hardware emulation offers high speed along with good debugging ability, but it is extremely expensive. Software simulation provides 100\% signal visibility, low cost and ease of debugging, but is the least efficient. Despite this drawback, software simulation remains the most commonly used method for RTL simulation~\cite{khronos}.


With the increasing scale of modern digital circuits, the simulation speed of the software approach decreases dramatically, and becomes the bottleneck in verification~\cite{eda17}. Table~\ref{table:CPU scale} presents the simulation speed of processors at different scales using Verilator~\cite{verilator}, a widely used open-source simulator. In this table, ``IR node'' refers to nodes in the RTL graph (including registers and logic units), and ``IR edge'' represents the number of connections between nodes, together indicating the complexity of each processor. As shown, the simulation speed for large-scale designs is typically limited to only a few kHz. Therefore, accelerating RTL simulation would significantly benefit the workflow for large-scale designs.

\begin{table}[htbp]
\caption{Verilator (single thread) simulation speed of booting Linux on different processors. The host CPU is Intel i9-9900K. stuCore is designed by undergraduate student.}
\label{table:CPU scale}
\begin{center}
\begin{tabular}{|c|c|c|c|c|}
\hline
Name & Configuration & IR node & IR edge & Speed \\
\hline
stuCore &  \makecell{In-Order\\ Single-Issue} & 9933 & 17369 & $\sim$ 900KHz \\
\hline
Rocket\cite{rocket} & \makecell{In-Order\\ Single-Issue} & 234807 & 293164 & $\sim$ 30KHz \\
\hline
BOOM\cite{boom} & \makecell{Out-of-Order\\ Triple-Issue} & 571038 & 827619 & $\sim$ 9KHz \\
\hline
XiangShan\cite{xiangshan} & \makecell{Out-of-Order\\ Six-Issue} & 6218427 & 9007005 & $\sim$ 0.9KHz \\
\hline
\end{tabular}
\end{center}
\end{table}


However, it is challenging to accelerate RTL simulation for large-scale designs. A software RTL simulator must simulate the signal propagation across all nodes in the circuit by computation. The heavy computation is the primary cause of the slow speed of RTL simulation. ESSENT~\cite{essent} proposes an essential signal simulation approach that significantly accelerates RTL simulation. We aim to further explore opportunities to reduce computation demands.


To identify the sources of computation, we further analyze the essential signal simulation approach. Listing~\ref{code:essent} shows the behavior of simulating a single cycle. Each node is associated with an {\ttfamily active} bit. During each cycle, the simulator traverses all nodes in the graph. If the {\ttfamily active} bit of a node is not set (Line 7), its value can be reused and the evaluation is skipped to reduce computation. Otherwise, the value of this node needs to be evaluated (Line 3). If the value changes after evaluation, its successors will be activated (Line 4-5).

\begin{minipage}[b]{0.4\columnwidth}
\begin{lstlisting}
eval1():
  update new value
cycle():
  eval1()
  eval2()
\end{lstlisting}
\captionsetup{type=listing} 
\captionof{lstlisting}{Simulation framework in Verilator.} 
\label{code:verilator}
\end{minipage}
\hfill
\begin{minipage}[b]{0.45\columnwidth}
\begin{lstlisting}[numbers=left]
eval1():
  save old value
  update new value
  if (old != new)
    active[succ] = true
cycle():
  if (active[1]) eval1()
  if (active[2]) eval2()
\end{lstlisting}
\captionsetup{type=listing} 
\captionof{lstlisting}{Simulation framework in ESSENT.}
\label{code:essent}
\end{minipage}

From the analysis above, the sources of computation can be concluded to four factors: \ding{182} the accessing of {\ttfamily active} bits, including checking and updating, \ding{183} the value evaluation in each node, \ding{184} the total number of nodes, \ding{185} the activity factor, which is the ratio of active nodes to the total number of nodes.


This paper proposes a novel RTL simulator, GSIM \footnote{http://github.com/OpenXiangShan/gsim} , which aims to reduce the overhead of the four factors mentioned above through three levels of granularity, including supernode level, node level and bit level.


\ding{182} \textbf{At the supernode level}, one challenge is to determine which nodes should be grouped into a supernode. A supernode is a set of nodes that are activated simultaneously, with the aim of reducing the activation overhead, but it may increase the activity factor. To balance this trade-off, we propose a novel partitioning algorithm to group nodes with strong correlations. This approach reduces the activation overhead while maintaining a low activity factor.


\ding{183} \textbf{At the node level}, GSIM employs several techniques. First, GSIM eliminates redundant nodes and optimizes expressions during value evaluation by data flow analysis. Second, we observe that although inlining the value evaluation of one node into another helps to reduce the number of nodes, it may introduce additional computation. To determine whether inlining is beneficial, GSIM constructs a cost model and selects the choice with lower cost. Finally, GSIM moves reset handling to the slow path, which reduces the number of checking reset signals by orders of magnitude.


\ding{184} \textbf{At the bit level}, we find that bits within each node may not be accessed simultaneously. By applying data flow analysis to each bit and splitting nodes according to their accessing patterns, GSIM can further reduce the activity factor.


GSIM successfully simulates XiangShan~\cite{xiangshan}, the state-of-the-art open-source RISC-V processor. When booting Linux on XiangShan, GSIM achieves a 7.34x speedup compared to Verilator. When running CoreMark~\cite{coremark} on Rocket~\cite{rocket}, GSIM reaches 19.94x of Verilator, outperforming the best between ESSENT~\cite{essent} and Arcillator~\cite{arcilator} by 2.52x.

\section{Background and Analysis}

\subsection{RTL Simulation}


To perform RTL simulation, the input design is represented as a directed graph. In this graph, each node corresponds to a register or logic unit, and each edge represents the propagation of signals between nodes. During simulation, the circuit's behavior is modeled as evaluation of signal values that propagate between nodes along the edges. The work of determining the evaluation order of nodes is called scheduling.


RTL simulation can be categorized into two types based on the scheduling policy~\cite{eda09}. In event-driven simulation, nodes are dynamically scheduled at runtime, with signal changes propagated as events. When the value of a node changes, an event is generated and sent to all its successors. This approach can flexibly model arbitrary delays, and is widely used in post-synthesis gate-level simulation. However, frequent scheduling in complex designs incurs high overhead. Commercial simulators like VCS~\cite{vcs} employ the event-driven simulation model.


In contrast, for full-cycle simulation, nodes are statically scheduled at compile time and evaluated in a fixed order at runtime. This eliminates the overhead of runtime scheduling. Although the minimum granularity of simulation is one clock cycle, it still suffices to perform functional verification for large-scale designs like processors. The open-source simulator Verilator~\cite{verilator} employs full-cycle simulation. Generally, full-cycle simulation is faster than event-driven simulation~\cite{dedup}.


In a basic implementation of full-cycle simulation, to build a directed acyclic graph (DAG), registers are often split into two nodes (one for reading and one for writing) to break cycles in the graph. Then a topological sort is performed on this graph. In each cycle, all nodes are evaluated in the topological order, as shown in Listing~\ref{code:verilator}. Verilator employs this approach.


To further improve the efficiency of full-cycle simulation, ESSENT~\cite{essent} introduces an essential signal simulation approach. This approach is effective, because only a few nodes have their value changed in a cycle. Listing~\ref{code:essent} shows the implementation. Each node is associated with an {\ttfamily active} bit. During a cycle, all {\ttfamily active} bits are examined. A node is evaluated only when its {\ttfamily active} bit is set. If the value of the node changes after evaluation, the active bits of its successors will be set to indicate the need of re-evaluation.

\subsection{Modeling and Analysis}



Inspired by ESSENT, we further explore optimization potential by identifying the sources of computation. According to the simulation framework in Listing~\ref{code:essent}, the simulation overhead per cycle $T$ can be modeled as $T = ((E + A_{succ}) * af + A_{exam}) * N$, where $E$ is the computation overhead of value evaluation in a node, $A_{succ}$ is the overhead of activating successors, $af$ is the activity factor, $A_{exam}$ is the overhead of examining the {\ttfamily active} bit, and $N$ is the number of nodes.


$E$ represents the computation required to evaluate the value of a node based on its predecessors. It contains various operations, including arithmetic, shifting, bitwise, and conditional operations.
$A_{succ}$ and $A_{exam}$ are introduced by the essential signal simulation approach. Each node requires an additional examination before evaluation and an activation step after evaluation. The former adds a branch instruction, while the latter involves a branch and several memory access instructions.
$N$ and $af$ are determined by the circuit design. They together indicate the number of nodes with the {\ttfamily active} bit set in a cycle, which require evaluation. As the complexity of the circuit increases, $N$ may grow a lot, but $af$ remains low in most designs. According to our evaluation, $af$ is about 4.61\% when running CoreMark on XiangShan.

Although $E$, $A_{succ}$, $A_{exam}$ and $af$ seem negligible, their impact on $T$ is amplified by a large $N$. Therefore, reducing $N$ can significantly improve simulation performance. Additionally, small improvements on the other factors can still yield substantial performance gains. For example, during the simulation of Xiangshan, 82.26\% of all executed branch instructions are dedicated to examining the {\ttfamily active} bits. Therefore, it is important to further optimize these factors.

\section{GSIM Design}


To accelerate RTL simulation, GSIM employs a series of optimization techniques targeting the factors above at three levels: super-node level, node level, and bit level. Note that some techniques may improve one factor while degrading another. In such cases, GSIM balances the trade-offs among these factors with corresponding strategies.

\subsection{Supernode Level Optimization}


Checking the {\ttfamily active} bit of each node separately will incur high overhead for $A_{exam}$ (Listing~\ref{code:essent}). To reduce $A_{exam}$, nodes are typically grouped into supernodes~\cite{essent}. Each supernode is associated with a single {\ttfamily active} bit. Nodes inside a supernode share an {\ttfamily active} bit and are evaluated together (Listing~\ref{code:supernode}). As a result, activating any node within a supernode leads to the evaluation of all nodes in it. We define the size of a supernode to be the number of nodes inside it. A supernode with larger size helps to reduce $A_{exam}$, but it may increase $af$.

\begin{figure}[htbp]
\begin{minipage}[b]{0.49\columnwidth}
\begin{lstlisting}
eval_super1():
  eval1()
  eval2()
cycle():
  if (active[1])
    eval_super1()
\end{lstlisting}
\captionsetup{type=listing} 
\captionof{lstlisting}{Supernode level activation.} 
\label{code:supernode}
\end{minipage}
\hfill
\begin{minipage}[b]{0.49\columnwidth}
\begin{lstlisting}[escapechar=$]
$\colorbox{lightgray}{if (active[7:0] != 0)}$
  if (active[1])
    eval_super1()
  if (active[2])
    eval_super2()
  ...
\end{lstlisting}
\captionsetup{type=listing} 
\captionof{lstlisting}{Check multiple bits with a single condition.}
\label{code:active flag simd}
\end{minipage}
\end{figure}

\begin{figure}[htbp]
\centering\includegraphics[width=0.85\columnwidth]{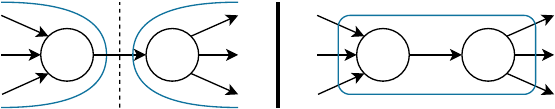}
\caption{Traditional partitioning algorithms fail to group nodes connected with few edges into the same partitions.}
\label{fig:bad partition}
\end{figure}

To keep $af$ low, GSIM tries to group nodes that are likely to be activated simultaneously into the same supernode. We observe that nodes in close proximity in the RTL graph often exhibit strong correlations and tend to be activated together. To find such nodes, a traditional graph partitioning algorithm may be a solution, such as Kernighan’s Algorithm~\cite{partition}. However, the goal of the traditional algorithm usually conflicts with building supernodes. This is because it tries to minimize the number of cut edges between partitions. It tends to divide nodes connected with a small number of edges into different partitions (the left of Figure~\ref{fig:bad partition}). However, such nodes may be activated together during simulation, and they should be grouped into the same supernodes (the right of Figure~\ref{fig:bad partition}).

To address this challenge, GSIM employs an enhanced version of Kernighan’s Algorithm. Specifically, the enhanced algorithm initially groups nodes with strong correlations based on certain rules, protecting them from being divided into different supernodes in the subsequent steps. These rules are derived from the following observations: \ding{182} A node with an out-degree of 1 is typically activated along with its successor. \ding{183} A node with an in-degree of 1 is usually activated when its predecessor is activated. \ding{184} Siblings with the same predecessors are always activated simultaneously. Subsequently, GSIM adopts the original Kernighan’s Algorithm to construct supernodes. The maximum size of a supernode can be controlled by a parameter in the algorithm.


Additionally, to take advantage of the low $af$, GSIM speculatively assumes multiple {\ttfamily active} bits are clear, and uses a single condition to examine them on the fast path (Line 1 in Listing~\ref{code:active flag simd}). If the condition holds, GSIM can skip the individual examination of each {\ttfamily active} bit to reduce $A_{exam}$.

\subsection{Node Level Optimization}



\begin{figure}[htbp]
\centering\includegraphics[width=\columnwidth]{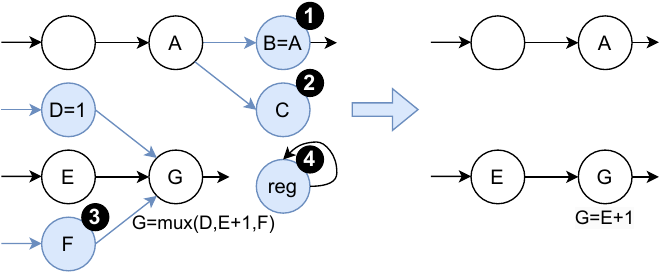}
\caption{Examples of eliminating redundant nodes.}
\label{fig:reduce node num}
\end{figure}

\textbf{Redundant node elimination}.
To reduce $N$, we eliminate the following types of redundant nodes (Figure~\ref{fig:reduce node num}) by applying data flow analysis.
\ding{182} \textbf{Alias Nodes} are duplicate nodes representing the same signal.
\ding{183} \textbf{Dead Nodes} are nodes not used by other nodes.
\ding{184} \textbf{Shorted Nodes} are nodes not selected due to another signal. 
\ding{185} \textbf{Unused Registers} are registers not used by other nodes, but they may be self-updated. 


\begin{figure}[htbp]
\centering\includegraphics[width=0.9\columnwidth]{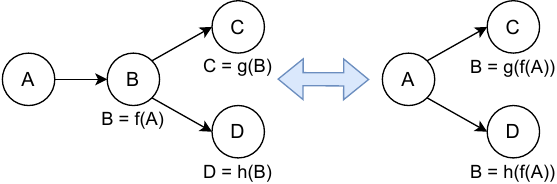}
\caption{Transform between node inline and extraction.}
\label{fig:merge node}
\end{figure}

\textbf{Node inline and extraction}.
For non-redundant nodes, GSIM considers whether to inline the evaluation of one node into its successors. This is a trade-off between $N$ and $E$. Extracting common parts of evaluation can reduce $E$ by reusing the value stored in an extra intermediate node. For example, in the left part of Figure~\ref{fig:merge node}, node B stores the result of f(A), and its successors (C and D) can directly use the value in B without re-evaluating f(A). Conversely, reducing $N$ may increase $E$. As shown in the right part of Figure~\ref{fig:merge node}, nodes C and D must re-evaluate f(A) without the intermediate node.

To address the trade-off above, GSIM models the cost of each decision and chooses the best one. Specifically, GSIM estimates the evaluation overhead of f(A), $cost(\mathrm{ f(A)})$, and the overhead of introducing a new node, $cost_{node}$, in terms of the number of operators involved, and the reference count of f(A), $\#\mathrm{f(A)}$. If $cost(\mathrm{f(A)}) \times \#\mathrm{f(A)} > cost(\mathrm{f(A)}) + cost_{node}$, f(A) is extracted as an extra node to reduce $E$. Otherwise, f(A) is inlined to reduce $N$.



\textbf{Activation overhead optimization}.
To optimize $A_{succ}$, ESSENT replaces the condition checking (Line 4 and 5 in Listing~\ref{code:essent}) with logical operations to relieve the pressure of the host branch predictor~\cite{essent}. However, we find that it may cause performance degradation when there are too many successors to activate, since this technique introduces extra operations for each successor to activate. To handle this trade-off, GSIM models the cost of each decision and chooses the optimal one, which is similar to the approach mentioned above.


\textbf{Reset handling optimization}.
During the evaluation of each register with reset port, the reset signal must be checked every cycle to determine the value to be written (Listing~\ref{code:reset before opt}). However, the number of reset signals is usually small, and they are hardly valid during simulation. Based on this observation, GSIM first speculatively assumes reset does not happen, and updates the register with the value evaluated based on its predecessors (Listing~\ref{code:reset opt}). This can remove the checking of reset signals from the fast path during evaluation, which helps to reduce $E$. After that, the reset signals are checked at the end of every cycle. In this way, GSIM can reduce the number of checking reset signals by orders of magnitude, from the number of registers with reset port to the number of reset signals in the design.

\begin{figure}[htbp]
\begin{minipage}[b]{0.49\columnwidth}
\begin{lstlisting}
eval1():
  reg1 = reset ? 0 : new_val
  ...
cycle():
  if (active[1]) ...
  ...
\end{lstlisting}
\captionsetup{type=listing} 
\captionof{lstlisting}{Before reset optimization.}
\label{code:reset before opt}
\end{minipage}
\hfill
\begin{minipage}[b]{0.49\columnwidth}
\begin{lstlisting}[escapechar=$]
eval1():
$\colorbox{lightgray}{  reg1 = new\_val}$
  ...
check_reset():
  if (reset)
    reg1 = 0
    reg2 = init_val2
    ...
    activate_all_succ()
\end{lstlisting}
\captionsetup{type=listing} 
\captionof{lstlisting}{After reset optimization.}
\label{code:reset opt}
\end{minipage}
\end{figure}


\textbf{Expression simplification}.
GSIM employs various techniques to optimize expressions during evaluation, such as constant propagation and recognition of complex patterns. For example, it is common to generate a one-hot signal and check each bit of it in RTL design. This may result in two nodes: one to evaluate {\ttfamily B = 1 << A}, and the other to evaluate {\ttfamily C = bits(B, k, k)}, where {\ttfamily k} is a constant. GSIM detects this pattern and optimizes it to  {\ttfamily C = (A == k)}.




\subsection{Bit Level Optimization}

\begin{figure}[htbp]
\centering\includegraphics[width=\columnwidth]{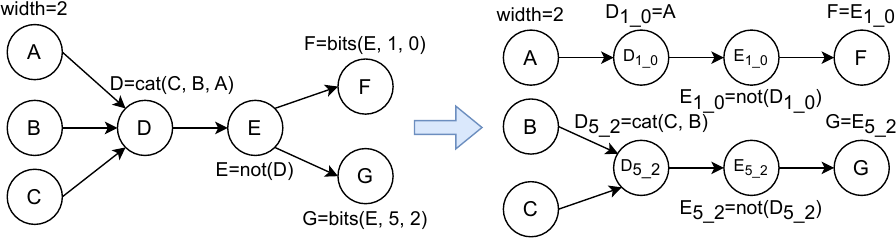}
\caption{Example of splitting nodes at the bit level.}
\label{fig:split bit}
\end{figure}

An activation is unnecessary if the value of the node remains the same after evaluation. We find that in many long signals, only certain bits change each cycle. If a successor only uses the unchanged bits, activating it will be unnecessary. As shown in the left part of Figure~\ref{fig:split bit}, if node A changes while B and C are inactive, both D and E will change, leading to the activation of F and G. However, the value of G remains unchanged, since it does not depend on A. Therefore, the activation of G is unnecessary. According to our statistics, among nodes with multiple bits in XiangShan, 23.7\% of them are composed of concatenations from other nodes, and 23.2\% of the references to these nodes only access a subset of their bits.

To address this issue, GSIM adopts a bit-level node splitting policy. Specifically, GSIM applies data flow analysis to each bit. For a node $P_0$, if there exists a bit set $S$ in $P_0$ and a path $P_0P_1\dots P_n$, where $P_n$ does not depend on $\overline{S}$ (the complementary bit set of $S$ in $P_0$), GSIM can split all nodes on the path according to $S$ and $\overline{S}$. In this way, we can remove the unnecessary activation of node $P_n$ when only bits in $\overline{S}$ change, thereby reducing $af$. As shown in the right part of Figure~\ref{fig:split bit}, node A will 
no longer trigger the activation of G.

Note that although splitting nodes may increase $N$, we can apply node level optimization techniques mentioned above to the generated nodes.

\subsection{Implementation Details}


GSIM accepts circuits in Firrtl~\cite{firrtl}, an intermediate representation for hardware that various HDL can be converted into. GSIM includes a Firrtl parser that converts the input design into an abstract syntax tree (AST) and further transforms it into a graph. Most optimizations are performed on this graph. After optimization, GSIM emits C++ simulation code. Additionally, GSIM allows users to adjust the maximum size of a supernode from the command line.

\section{Evaluation}

\subsection{Evaluation Setup}



We compare GSIM with the following simulators. \ding{182} \textbf{Verilator}~\cite{verilator} is the most widely used open-source Verilog simulator. We use version v5.026 with {\ttfamily -O3} flag to generate C++ files. It also supports multi-threaded simulation. \ding{183} \textbf{Arcilator}~\cite{arcilator} is an MLIR~\cite{mlir} simulator developed as part of the CIRCT~\cite{circt} project. \ding{184} \textbf{ESSENT} is a single-threaded Firrtl~\cite{firrtl} simulator, with {\ttfamily -O3} flag to generate C++ files.

\begin{figure}[htbp]
\centering\includegraphics[width=\columnwidth]{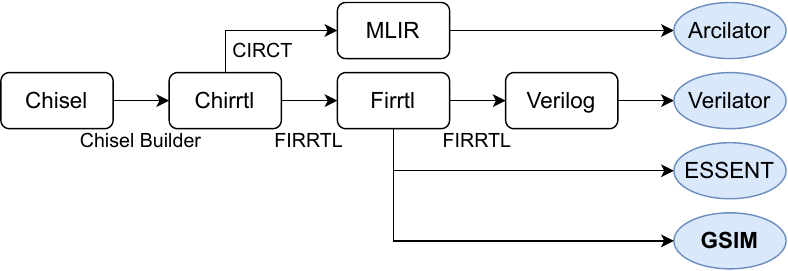}
\caption{The input of different simulators.}
\label{fig:run flow}
\end{figure}

We evaluate the simulators on processors in Table~\ref{table:CPU scale}. Table~\ref{table:processor version} shows their versions. All processors are mainly developed in Chisel~\cite{chisel}. For a few Verilog modules, we replace them with Chisel implementations. As shown in Figure~\ref{fig:run flow}, each design is first converted into Chirrtl IR, and then translated into Firrtl, Verilog, and MLIR files using the FIRRTL~\cite{firrtl} compiler or CIRCT~\cite{circt} compiler. ESSENT and GSIM accept Firrtl as input, Verilator accepts Verilog, and Arcilator accepts MLIR.


\begin{table}[htbp]
\caption{The version of the processors.}
\label{table:processor version}
\begin{center}
\begin{tabular}{|c|c|c|}
\hline
          & commit hash (truncated)  & commit date \\
\hline
Rocket    & e3773366a5c473b6b45107f037e3130f4  & Sep 26, 2023 \\
\hline
BOOM      & 9459af0c1f6847f8411622dac770ac78f  & Oct 17, 2023 \\
\hline
XiangShan & a42a7ffe5e4be903c75d53cf243761ce7  & Mar 1, 2024 \\
\hline
\end{tabular}
\end{center}
\end{table}

We choose CoreMark~\cite{coremark} and Linux as the software workloads. The former exhibits hot spots, while the latter does not. The host configuration is Intel 8-core i9-9900K CPU at 3.60GHz with 64GB of RAM. The host OS is Debian 12.

\subsection{Overall Performance}

\begin{figure}[htbp]
\includegraphics[width=\columnwidth]{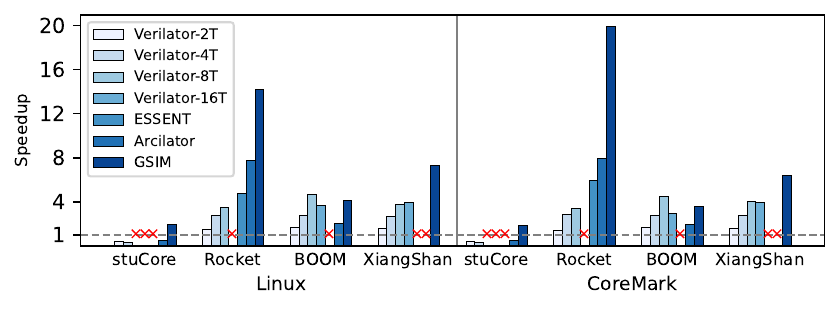}
\caption{Overall performance. The speedup is normalized to the single thread version of Verilator. `2T' means simulating with 2 threads. `$\times$' means failing to perform the simulation.}
\label{fig:overall perf}
\end{figure}

Figure~\ref{fig:overall perf} shows the performance of different simulators on various processors and software workloads. Since Verilator is the most widely used open-source RTL simulator and successfully simulates all selected processors, we use Verilator as the baseline and normalize the other results to it.

In terms of correctness, aside from Verilator, GSIM is the only simulator that can successfully simulate XiangShan. ESSENT and Arcilator fail to simulate some of the designs due to errors during emitting C++ files or wrong simulation results. For performance, GSIM achieves speedup of 7.34x for booting Linux on XiangShan, and 19.94x for running CoreMark on Rocket. We also evaluate Verilator with multi-threading. GSIM significantly outperforms multi-threading Verilator on most of the designs.

\subsection{Running SPEC CPU2006}



\begin{figure}[htbp]
\includegraphics[width=\columnwidth]{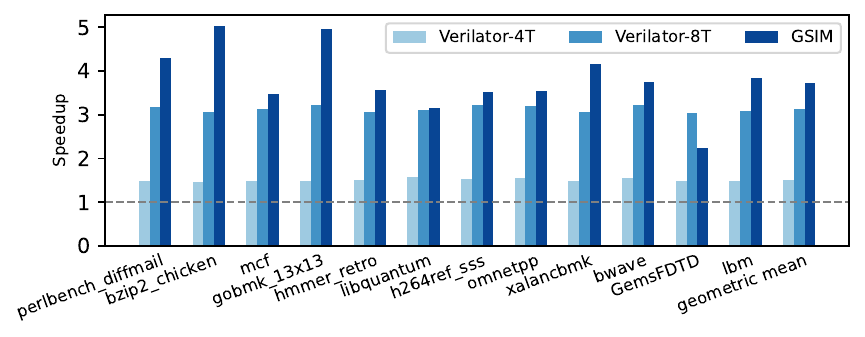}
\caption{SPEC CPU2006 performance on XiangShan. The speedup is normalized to the single thread version of Verilator.}
\label{fig:speccpu perf}
\end{figure}

In CPU simulation, checkpoints have long been used to accelerate performance evaluation~\cite{xiangshan}~\cite{checkpoint}. We sample SPEC CPU2006~\cite{speccpu} using SimPoint~\cite{simpoint} and extract 40M-instruction segments to create checkpoints. The selected benchmarks cover various types of workloads, including streaming and irregular memory access, integer and floating-point computation, branch-intensive operations, and instruction-cache-intensive~\cite{spec2006-analysis}. We run these checkpoints on XiangShan using both Verilator and GSIM. As shown in Figure~\ref{fig:speccpu perf}, GSIM is 3.72x faster than single-threaded Verilator across all checkpoints on average, and achieves an average 1.18x of 8-threaded Verilator.

\subsection{Performance Breakdown}

\begin{figure}[htbp]
\includegraphics[width=\columnwidth]{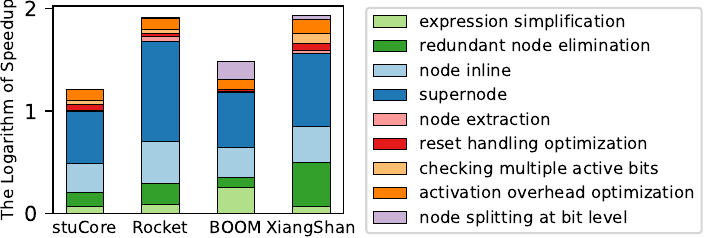}
\caption{Performance breakdown for each technique.}
\label{fig:perf breakdown}
\end{figure}

To evaluate the contribution of each optimization technique adopted in GSIM, we apply all techniques incrementally, starting with a baseline with no optimizations. In this way, we get a series of performance data $P_0, P_1, \dots, P_n$. In Figure~\ref{fig:perf breakdown}, the height of the bar for the $i$th technique is calculated as $\log_{10}(P_{i}/P_{i-1})$. As shown in the figure, introducing supernode significantly improves performance for all designs. Splitting nodes at bit level delivers substantial improvement for BOOM and notable gains for XiangShan, but has little impact on stuCore and Rocket.

\subsection{Selecting the Maximum Size of Supernode}

\begin{figure}[htbp]
\includegraphics[width=\columnwidth]{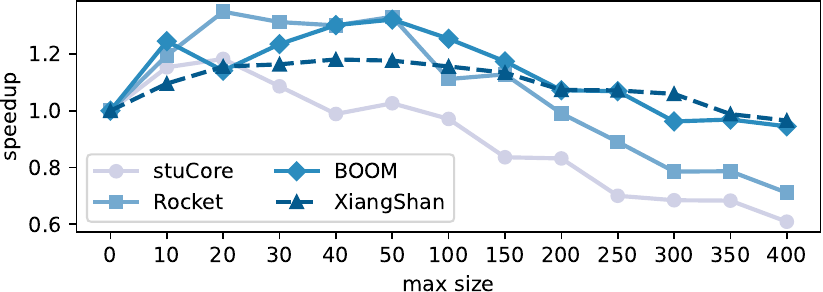}
\caption{Performance with respect to maximum supernode size.}
\label{fig:partition threshold}
\end{figure}

We also investigated the relationship between the maximum size of supernode and the simulation performance. As mentioned above, a larger size of supernode reduces $A_{exam}$ but increases $af$. The optimal size depends on the circuit design. To find the optimal size, we try different sizes with all other optimization techniques enabled. For the designs we select, the optimal size ranges from 20 to 50, as shown in Figure~\ref{fig:partition threshold}.

\subsection{Different Algorithms to Build Supernode}


We compare the enhanced partitioning algorithm used in GSIM against the original Kernighan's Algorithm and ESSENT's partitioning algorithm (noted as MFFC-based). These algorithms are implemented on GSIM with all other optimization techniques disabled to show their effectiveness. They are evaluated by running CoreMark on BOOM under their own optimal parameters. We measure the number of supernodes, the number of times to active a successor, and the number of active node. These metrics can reflect $A_{exam}$, $A_{succ}$ and $E$ respectively. As shown in Table~\ref{table:partition}, our algorithm performs the best in terms of the simulation speed, by achieving a better balance among the three factors above.

\begin{table}[htbp]
\caption{Comparison of different partitioning algorithms by running CoreMark on BOOM.}
\label{table:partition}
\begin{center}
\begin{tabular}{|c|c|c|c|c|c|}
\hline
                & \makecell{partition \\ time (s)} & supernode & \makecell{activation \\ times} &\makecell{active \\ node} & \makecell{speed \\ (Hz)}\\
\hline
None            & -- & 536721 & 64044  & 42398 & 913 \\
\hline
Kernighan       & 1.3 & 21769&  13104  & 93327  & 4522 \\
\hline
MFFC-based     & 88.9 & 36478 & 7310  & 124202 & 6802 \\
\hline
GSIM           & 1.6  & 14261 & 8169  & 114255 & 8184 \\
\hline
\end{tabular}
\end{center}
\end{table}

\subsection{Resource Usage}


For the emission time, we measure the time required to emit simulation files.
As shown in Table~\ref{table:resources}, although GSIM outperforms Verilator, GSIM still delivers comparable emission time to Verilator. ESSENT requires more time to generate C++ files. One reason may be the slow execution of Scala programs. Arcilator requires more than 100GB memory to emit code for BOOM, which leads to accessing swap with a very long emission time. It even fails to emit code for XiangShan due to out-of-memory.

For the code size, we measure the size of the {\ttfamily .text} section of the binary reported by {\ttfamily readelf}.
As shown in Table~\ref{table:resources}, GSIM emits the least amount of code among all simulators, as it employs more optimization techniques.

For the data size, we use {\ttfamily sizeof} to obtain the size of the top-level design class in C++ files, whose members contain all variables used for simulation, and we exclude the array used to simulate the main memory (128MB). As shown in Table~\ref{table:resources}, GSIM generates slightly larger data than Verilator. 


\begin{table}[htbp]
\caption{Comparison of resources across various designs. Instances labeled with `*' mean failing to run.}
\label{table:resources}
\begin{center}
\begin{tabular}{|c|c|c|c|c|}
\hline
Design    & Simulator  & \makecell{Emission \\ Time(s)} & \makecell{Code \\Size(B)}  & \makecell{Data \\Size(B)} \\
\hline
\multirow{4}*{stuCore}   & Verilator  & 1.1     & 394K     & 15K   \\
\cline{2-5}              & *ESSENT    & 7.7     & 659K     & 21K   \\
\cline{2-5}              & Arcilator  & 0.4     & 131K     & 13K  \\
\cline{2-5}              & GSIM      & 0.4     & 133K     & 16K \\
\hline
\multirow{4}*{Rocket}    & Verilator  & 5.8     & 2.9M     & 62K   \\
\cline{2-5}              & ESSENT     & 37.8    & 1.6M     & 107K   \\
\cline{2-5}              & Arcilator  & 2.9     & 305K     & 49K  \\
\cline{2-5}              & GSIM      & 9.8     & 1.4M     & 72K \\
\hline
\multirow{4}*{BOOM}      & Verilator  & 22.9    & 7.7M     & 954K   \\
\cline{2-5}              & *ESSENT    & -       & -        & -   \\
\cline{2-5}              & Arcilator  & 4194.6  & 2.8M     & 799K  \\
\cline{2-5}              & GSIM      & 31.3    & 4.4M     & 976K \\
\hline
\multirow{4}*{XiangShan} & Verilator  & 374.6   & 40.7M    & 22.2M   \\
\cline{2-5}              & *ESSENT    & -       & -        & -   \\
\cline{2-5}              & *Arcilator & -       & -        & -  \\
\cline{2-5}              & GSIM      & 389.1   & 25.4M    & 22.3M \\
\hline
\end{tabular}
\end{center}
\end{table}

\section{Related Work}




Verilator~\cite{verilator} is a widely used open-source, full-cycle RTL simulator. It supports multi-threading by partitioning the RTL design. Arcilator~\cite{arcilator}, an RTL simulator based on CIRCT, applies optimizations at various IR levels. Both of them evaluate all signals in the circuit every cycle and improve simulation speed through various expression optimization techniques. In contrast, GSIM skips the evaluation of a signal when it is inactive.
Khronos~\cite{khronos} employs cross-cycle optimizations to reduce memory access overhead during simulation, which performs well in circuits with many pipeline registers. However, GSIM targets more general designs.
ESSENT~\cite{essent} leverages the low activity factor in circuits, partitioning the design and filtering out inactive regions to efficiently reduce the computation overhead. GSIM further explores the potential of accelerating RTL simulation. It reduces the computation overhead from four factors with optimization techniques across three levels.


RepCut~\cite{repcut} is a multi-threaded RTL simulator based on ESSENT. It reduces inter-thread synchronization overhead by introducing redundant computation. \cite{dedup} proposes a coarse-grained circuit deduplication method to improve throughput for a batch of simulation tasks. GSIM focuses on improving the performance of single-threaded simulation under a single task. These works are orthogonal to GSIM and provide directions for future work.


\section{Conclusion}



This paper proposes GSIM, a novel RTL simulator based on the essential signal simulation approach.
GSIM employs a series of optimization techniques at the supernode level, node level, and bit level. GSIM successfully simulates XiangShan, the state-of-the-art open-source RISC-V processor. Compared to Verilator, GSIM achieves a 7.34x speedup when booting Linux on XiangShan, and a 19.94x speedup when running CoreMark on Rocket.
 
\ifx\cameraready\undefined
\else

\section*{acknowledgement}
The authors would like to thank the anonymous reviewers
for their valuable feedback and comments.
This work is supported by the National Key R\&D Program of China (2022YFB4500403).


\fi


\bibliographystyle{./IEEEtran}
\bibliography{reference}
\end{document}